\documentclass[twocolumn,aps,pra,showpacs,amsmath,amssymb,superscriptaddress]{revtex4-1}
\preprint{APS/123-QED}

\usepackage{graphicx}
\usepackage{dcolumn}
\usepackage{bm}
\usepackage{hyperref}

\begin{document}

\title{Collapse and revival of the monopole mode of a degenerate Bose gas in an isotropic harmonic trap}

\author{C.~J.~E.~Straatsma}
\email{cameron.straatsma@colorado.edu}
\affiliation{JILA and Department of Electrical, Computer, and Energy Engineering, University of Colorado, Boulder, Colorado 80309-0440, USA}

\author{V.~E.~Colussi}
\affiliation{JILA, NIST, and Department of Physics, University of Colorado, Boulder, Colorado 80309-0440, USA}

\author{M.~J.~Davis}
\affiliation{School of Mathematics and Physics, University of Queensland, Brisbane, Queensland 4072, Australia}
\affiliation{JILA, University of Colorado, Boulder, Colorado 80309-0440, USA}

\author{D.~S.~Lobser}
\altaffiliation{Current address: Sandia National Laboratories, Albuquerque, NM 87185-1086, USA.}
\affiliation{JILA, NIST, and Department of Physics, University of Colorado, Boulder, Colorado 80309-0440, USA}

\author{M.~J.~Holland}
\affiliation{JILA, NIST, and Department of Physics, University of Colorado, Boulder, Colorado 80309-0440, USA}

\author{D.~Z.~Anderson}
\affiliation{JILA, NIST, and Department of Physics, University of Colorado, Boulder, Colorado 80309-0440, USA}

\author{H.~J.~Lewandowski}
\affiliation{JILA, NIST, and Department of Physics, University of Colorado, Boulder, Colorado 80309-0440, USA}

\author{E.~A.~Cornell}
\affiliation{JILA, NIST, and Department of Physics, University of Colorado, Boulder, Colorado 80309-0440, USA}

\date{\today}

\begin{abstract}
We study the monopole (breathing) mode of a finite temperature Bose-Einstein condensate in an isotropic harmonic trap recently developed by Lobser \emph{et al.} [Nat.~Phys. \textbf{11}, 1009 (2015)].  We observe a nonexponential collapse of the amplitude of the condensate oscillation followed by a partial revival. This behavior is identified as being due to beating between two eigenmodes of the system, corresponding to in-phase and out-of-phase oscillations of the condensed and noncondensed fractions of the gas.  We perform finite temperature simulations of the system dynamics using the Zaremba-Nikuni-Griffin methodology [J.~Low Temp.~Phys. \textbf{116}, 277 (1999)], and find good agreement with the data, thus confirming the two mode description.
\end{abstract}

\maketitle

\section{Introduction}
In strongly interacting systems, the study of collective modes of a quantum many-body problem provides revealing information about the nature of the underlying Hamiltonian. In ultracold atomic gases, due to the fact that binary interactions between particles at low energy are well understood, the experimental measurement of collective modes provides a means of evaluating and potentially falsifying many-body theoretical methods used to describe these systems. Experiments probing the collective modes of ultracold gases were carried out shortly after the demonstration of Bose-Einstein condensation in a dilute atomic vapor~\cite{becform1,becform2}. In early experiments at JILA~\cite{colmodebec1} and MIT~\cite{colmodebec3}, the low-lying quadrupole modes of a nearly pure Bose-Einstein condensate (BEC) were excited, and the observed oscillation frequencies showed good agreement with the Bogoliubov spectrum~\cite{bogo,stringari}. Experiments were then conducted over a range of temperatures below the critical point, and temperature-dependent shifts in the oscillation frequencies and damping rates were observed~\cite{colmodebec2,colmodebec4}. At the time, existing theoretical models were unable to reproduce the experimental observations.

The observation of these unexplained temperature-dependent shifts motivated further exploration of collective-mode behavior at finite temperature, where experiments probed the interaction between the condensate and thermal component of the gas (i.e., noncondensate). Experiments on the scissors modes~\cite{scissors2} provided an alternative means of measurement of temperature-dependent shifts of the mode frequencies and damping rates through observation of the angle oscillations of the condensate and noncondensate, and reasonable agreement with existing theories was found. Furthermore, a study of the transverse breathing mode in an elongated harmonic trap~\cite{transbreath} found uncharacteristically small damping rates and observed that the mode frequency was quasi-independent of temperature.

In order to address the unexpected behavior of the experiments, models were initially developed to explain the anomalous temperature dependence of the quadrupole mode found in Ref.~\cite{colmodebec2}.  Early efforts assuming a static noncondensate were unable to reproduce the experimental results;  however, inclusion of the dynamics of the noncondensate lead to a consistent framework that matched the experiment.  Using a semi-classical coupled-modes model, Stoof, Bijlsma, and Al Khawaja~\cite{stoof,stoof2} described the coupled dynamics of the condensate and noncondensate in terms of in-phase and out-of-phase eigenmodes, which are collisionless analogs of first and second sound hydrodynamic modes~\cite{sound1,sound2,sound3}. They concluded that the anomalous behavior found in Ref.~\cite{colmodebec2} was the result of simultaneous excitation of both eigenmodes of the system.  Numerical simulation of the Zaremba-Nikuni-Griffin (ZNG) equations by Jackson and Zaremba~\cite{quadzng,zngscissors,zngtransbreath} confirmed this picture, and Morgan, Rusch, Hutchinson, and Burnett provided additional analysis in an extension of their previous work~\cite{morgan0,morgan01,morgan1,morgan2,morgan3}. These efforts highlighted the important role of the noncondensate dynamics in the behavior of collective modes at finite temperature.

Experiments to date have operated with anisotropic trapping geometries, which lead to an increased degree of complexity in the collective-mode spectrum. An isotropic harmonic trap simplifies the mode spectrum due to its spherical symmetry, and allows for a detailed comparison between experimental measurements and existing theoretical models. Furthermore,  a spherical trapping geometry eases the computational burden of sophisticated numerical studies such as the simulation of the ZNG equations. However, experiments in this regime face the technological hurdle of minimizing asphericities in the trapping potential, which to date has  prevented the study of collective modes in such a simplified geometry. Thus, fundamental comparisons between theoretical predictions for the collective-mode spectrum of condensates, as well as their frequency shifts and damping rates at finite temperature, have yet to be made.

In this paper, we present experimental measurements and analysis of the monopole mode of a finite temperature BEC confined in an isotropic harmonic trap. We observe a collapse and partial revival of the condensate oscillation, and compare these results to the predictions of finite temperature BEC models. We set the scene in Sec.~\ref{sec:collisionless} by providing an overview of the limiting cases for the collective modes of a BEC and ideal gas in an isotropic harmonic trap. This is followed by a theoretical analysis of the spectrum of coupled modes of the condensate and noncondensate at finite temperatures in Sec.~\ref{sec:cmodes_analysis}. This provides a framework for understanding the collapse and revival time scales observed in the experiment. In Sec.~\ref{sec:experiment} we provide a description of the experimental procedure and the main results of this paper. In Sec.~\ref{sec:zng} we investigate the damping observed experimentally through numerical simulations within the ZNG formalism. After discussing the results of the numerical simulations, we conclude in Sec.~\ref{sec:conclusion}. In Appendix~\ref{appendix:cfield} we describe efforts towards reproducing the experimental observations with classical field methods, and provide a comparison of those results to the ZNG simulations. Appendix~\ref{appendix:zng} provides technical details of the numerical solution of the ZNG equations for an isotropic trapping geometry.

\section{Collisionless Dynamics}\label{sec:collisionless}
Here, we provide an overview of the collisionless dynamics of a trapped Bose gas, beginning with a discussion of the limiting cases for collective modes in an isotropic trap. We then discuss the monopole mode of a finite temperature BEC through application of the semiclassical collisionless model from Ref.~\cite{stoof,stoof2} to a spherically symmetric trapping geometry, and show how the monopole mode response can be cast in terms of two eigenmodes of the system. This analysis provides a framework for understanding the collapse and revival behavior of the condensate oscillation observed in the experimental results.

\subsection{Collective modes in an isotropic harmonic trap}\label{sec:coll_modes}
In an isotropic harmonic trap, the collective modes of a Bose gas are well understood in two limits.  In the Thomas-Fermi (TF) limit at zero temperature the ratio of the kinetic to interaction energy is small when the number of atoms in the BEC is large; thus, the kinetic energy can be neglected. The collective-mode frequencies of a BEC in a three-dimensional isotropic harmonic trap in this limit can be estimated using a hydrodynamic approach~\cite{stringari}. The mode frequencies depend on the principle quantum number $n$ and angular quantum number $l$ according to
\begin{equation}\label{eq:spectrum}
\omega^2 = \omega_0^2\left(l+3n+2nl+2n^2\right),
\end{equation}
where $\omega_0$ is the harmonic trap frequency and $\omega$ is the frequency of the collective mode. For the spherically symmetric monopole, or breathing mode ($n = 1, l = 0$), the mean-square radius of the condensate oscillates at $\omega = \sqrt{5}\omega_0$, and the motion is undamped. Above the BEC critical temperature, $T_c$, mean-field effects can be neglected and the gas can be described by a classical Boltzmann equation. In this case, the mode oscillates at $\omega = 2\omega_0$ in both the collisionless and hydrodynamic regimes~\cite{odelinosc}, and the motion is undamped. In the collisionless regime individual atoms may undergo many oscillations before experiencing a collision while the hydrodynamic regime implies the gas is in local statistical equilibrium.

\subsection{Coupled-modes analysis}\label{sec:cmodes_analysis}
To obtain insight into the behavior of the monopole mode at finite temperature in an isotropic trap we apply a model previously developed by Bijlsma and Stoof~\cite{stoof}. This methodology introduces a dynamical scaling ansatz for the condensate and noncondensate that successfully reproduces the limiting cases of the monopole mode behavior described in the previous section. The condensate and noncondensate are described by a Gross-Pitaevskii equation (GPE) and a collisionless quantum Boltzmann equation (QBE), respectively, which are coupled by their mean-field interaction. The analysis here assumes a small amplitude perturbation of the system, and a linear response such that the effects of damping are absent.

In the following calculation, a scaling ansatz is made for the time evolution of the condensate density,
\begin{equation}
n_c(\textbf{r},t) = \frac{1}{\lambda^3}n_c^0\left(\frac{\textbf{r}}{\lambda}\right),
\end{equation}
and the Wigner distribution function of the noncondensate,
\begin{equation}\label{eq:scalingansatz}
f(\textbf{r},\textbf{p},t) = \frac{1}{\bar{\alpha}^{6}}f^0\left(\frac{\textbf{r}}{\alpha\bar{\alpha}},\frac{\alpha}{\bar{\alpha}}\left[\textbf{p}-\frac{m\dot{\alpha}}{\alpha}\right]\right),
\end{equation}
which are written in terms of a Gaussian density profile $n_c^0$ for the condensate and a saturated Bose-Einstein distribution $f^0$ for the noncondensate:
\begin{eqnarray}
n_c^0(\textbf{r}) & = & N_c\left(\frac{m\omega_0}{\pi\hbar}\right)^{1/2}e^{-\frac{m\omega_0}{\hbar}r^2}, \nonumber \\
f^0(\textbf{r},\textbf{p}) & = & \tilde{N}\left(\frac{\hbar\omega_0}{k_BT\zeta(3)}\right) \nonumber \\
& \times & \left[e^{\left(\frac{p^2}{2m}+\frac{1}{2}m\omega_0^2r^2\right)/k_BT}-1\right]^{-1}.
\end{eqnarray}
The number of atoms in the condensate and noncondensate are denoted by $N_c$ and $\tilde{N}$, respectively, and $\zeta(s)$ is the Riemann zeta function. The scaling parameters, $\lambda(t)$ and $\alpha(t)$, capture the oscillation of the widths of the two components, and the bar denotes the equilibrium value. Inserting the scaling ansatz into the GPE and QBE results in a set of coupled equations of motion for the condensate and noncondensate characteristic widths:
\begin{equation}\label{eq:coupledeq}
\ddot{{\bf u}}+\omega_0^2{\bf u}={\bf v}({\bf u}),
\end{equation}
where the vector ${\bf u}$ contains the scaling parameters
\begin{equation}
{\bf u}=\left(\begin{array}{c}
\lambda\\
\alpha\\
\end{array}
\right),
\end{equation}
and ${\bf v}({\bf u})$ is a nonlinear vector function describing the spreading of the cloud due to kinetic energy and the effects of nonlinear interactions (see Ref.~\cite{stoof} for details).

\begin{figure}[!]
\includegraphics[scale=1]{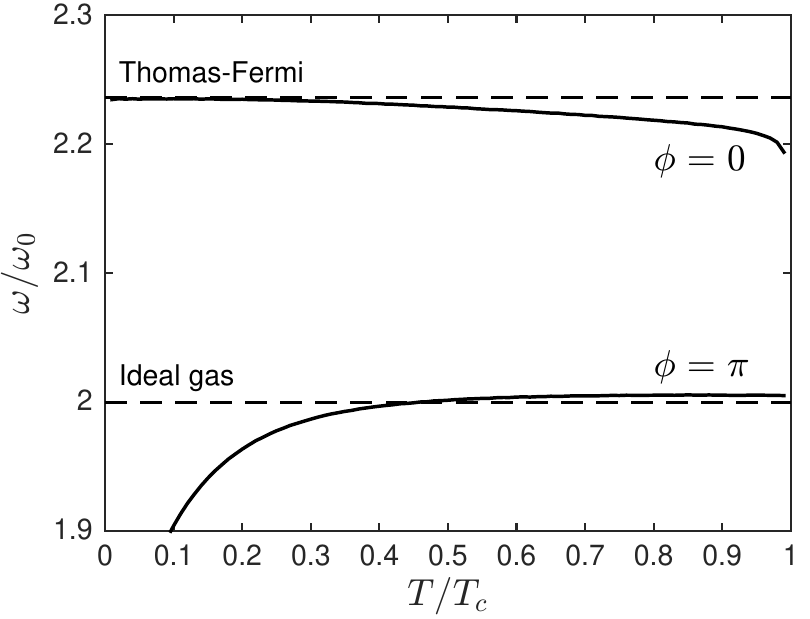}
\caption{\label{fig:freqcomp} Oscillation frequency of the in-phase ($\phi = 0$) and out-of-phase ($\phi = \pi$) modes as a function of temperature from the coupled-modes analysis (solid lines). The dashed lines represent the TF limit ($\sqrt{5}\omega_0$) and ideal gas limit ($2\omega_0$) for the monopole mode frequency of the condensate and noncondensate, respectively. Assumptions made in the coupled-modes analysis become invalid for $T\lesssim0.2~T_c$.}
\end{figure}

In the limit of a small amplitude oscillation, the total density of the system can be written as
\begin{equation}\label{eq:densitypert}
n(\textbf{r},t)=\bar{n}(\textbf{r})+\delta n(\text{r}) e^{i\omega t},
\end{equation}
where the perturbation is generated by modulating the trap frequency with amplitude $\epsilon$:
\begin{equation}\label{eq:trappert}
\omega_0(t)=\left(1+\epsilon e^{i\omega t}\right)\omega_0.
\end{equation}
In this limit Eq.~\eqref{eq:coupledeq} can be linearized:
\begin{equation}\label{eq:linearizedeq}
-\omega^2\delta{\bf u}+\omega_0^2\delta {\bf u}=\left[\nabla_{{\bf u}}{\bf v}\right]\big|_{\bar{\bf u}}\cdot\delta{\bf u}-2\epsilon\omega_0^2\bar{{\bf u}},
\end{equation}
and the eigenfrequencies $\omega_n$ and eigenmodes ${\bf u}^{(n)}$ of the homogeneous part of Eq.~\eqref{eq:linearizedeq} can be extracted with a solution of the form
\begin{equation}
\delta{\bf u}=2\epsilon\omega_0^2\sum_n\frac{{\bf u}^{(n)}\cdot\bar{\bf{u}}}{\omega^2-\omega_n^2}{\bf u}^{(n)}.
\end{equation}
From this solution, we find two eigenmodes that we refer to as the in-phase and out-of-phase modes of the system. The in-phase mode corresponds to the condensate and noncondensate monopole modes oscillating together with a phase difference of $\phi = 0$, and the out-of-phase mode corresponds to a phase difference of $\phi = \pi$. In Fig.~\ref{fig:freqcomp} the frequencies of the in-phase and out-of-phase modes as a function of temperature are shown.

\begin{figure}[!]
\includegraphics[scale=1]{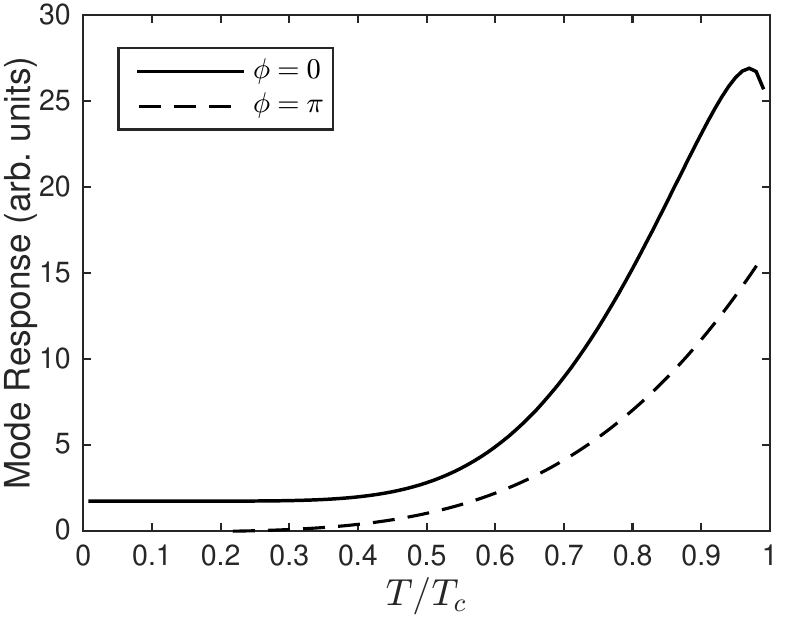}
\caption{\label{fig:residues} Magnitude of the response of the in-phase ($\phi = 0$) and out-of-phase ($\phi = \pi$) modes to a trap frequency perturbation as a function of temperature. The amplitude of the trap frequency modulation is $\epsilon = 0.01$.}
\end{figure}

Given the eigenmodes of the system, the time-averaged work done by a perturbation of the trap frequency can be used to characterize the response of the system:
\begin{equation}
W=\sum_n\frac{b_n}{\omega^2-\omega_n^2},
\end{equation}
where the $b_n$ are a measure of the magnitude that each eigenmode responds with when the system is perturbed. Figure~\ref{fig:residues} shows the $b_n$ as a function of temperature for the two modes discussed above.
For temperatures $T>0.2~T_c$ both modes of the system will be excited by a perturbation of the trap frequency.

The coupled-modes analysis suggests that the oscillation of a single component of the gas (e.g., condensate) is described by a superposition of two eigenmodes oscillating at slightly different frequencies. Therefore, we expect measurements of the condensate width as a function of time for temperatures $T>0.2~T_c$ to beat at a frequency corresponding to the frequency difference between the two eigenmodes, $\Delta\omega/\omega_0\sim0.2$--$0.25$ (see Fig.~\ref{fig:freqcomp}). Given this result, we present experimental observations of the monopole mode in an isotropic trap in the next section.

\section{Experiment}\label{sec:experiment}
The experimental system is a Bose gas of $^{87}\text{Rb}$ atoms cooled to quantum degeneracy via forced radio-frequency evaporation in a time-averaged, orbiting potential (TOP) trap~\cite{toptrap95}.  A standard TOP trap configuration results in an oblate harmonic trap with an aspect ratio of $\omega_z/\omega_r = \sqrt{8}$, where $\omega_z$ ($\omega_r$) is the axial (radial) trapping frequency.  Here, the overall harmonic confinement of the trap is reduced and the trap minimum is allowed to sag under the force of gravity. This causes the curvature of the magnetic field along the $z$ axis to decrease, which effectively decreases the ratio $\omega_z/\omega_r$. The end result is an isotropic harmonic trap with $\omega_0 \equiv \omega_r = \omega_z  = 2\pi\times (9.03(2)~\text{Hz})$ with a residual asphericity of less than $0.2\%$. This system was used in previous work to study the monopole mode of a Bose gas above the BEC critical temperature~\cite{breath2015}, and a detailed description of the apparatus can be found in Ref.~\cite{danthesis}.

We excite the monopole mode of the system below the BEC critical temperature in the range of approximately $0.75$--$0.9~T_c$.  The experimental procedure parallels that of Ref.~\cite{breath2015}---beginning from a system at equilibrium, the trap frequency is sinusoidally modulated at a driving frequency $\omega_D\approx2\pi\times18$--$19~\text{Hz}$ for four periods with an amplitude $\epsilon\approx0.1$:
\begin{equation}
\omega(t) = \left[1+\epsilon\sin{\left(\omega_Dt\right)}\right]\omega_0.
\end{equation}
After driving, we find that the peak TF radius of the condensate is $10$--$15\%$ larger than the equilibrium value for all of the experimental data sets. The system is then allowed to freely evolve in the static isotropic trap for a time $t$ before six nondestructive phase-contrast images record the integrated column density of the cloud at intervals of $10~\text{ms}$ or $17~\text{ms}$, sampling between $1$ and $1.5$ oscillation periods of the monopole mode. This experimental procedure is repeated between $2$--$4$ times for each $t$, and for times up to $t\approx1.5~\text{s}$.

Each phase-contrast image is analyzed using a 2D bimodal fit to the atomic column density. The fitting function is the sum of a Gaussian and integrated TF function~\cite{ketterle1999varenna}:
\begin{eqnarray}\label{eq:bimodfit}
n_{\text{col}}(x,z) & = & A_G\exp{\left[-\left(\frac{x-x_c}{\sigma_{G,x}}\right)^2-\left(\frac{z-z_c}{\sigma_{G,z}}\right)^2\right]} \nonumber \\
& + & A_{TF}\left[1-\left(\frac{x-x_c}{\sigma_{TF,x}}\right)^2-\left(\frac{z-z_c}{\sigma_{TF,z}}\right)^2\right]^{3/2} \nonumber \\
& + & C_{\text{col}},
\end{eqnarray}
where $A_G$ and $A_{TF}$ are the amplitudes of the Gaussian and TF functions, respectively, $x_c$ and $z_c$ are the center points of the cloud, $\sigma_{G,i}$ are the Gaussian widths, $\sigma_{TF,i}$ are the TF widths, and $C_{\text{col}}$ is a constant offset. Note that the TF function is defined to be zero if the argument in brackets is negative. 

The dynamics of the condensate monopole mode are captured by the spherically symmetric quantity
\begin{equation}\label{eq:AMc}
\sigma_M^2 = \left(\sigma_{TF,x}^2+\sigma_{TF,y}^2+\sigma_{TF,z}^2\right)/3.
\end{equation}
During the data runs for this experiment, images were consistently taken in the $xz$ plane. In earlier measurements described in Ref.~\cite{breath2015}, data were also taken along the $xy$ plane, but technical difficulties were encountered with the imaging system along this axis during the course of the experiments discussed here. However, the limited data available from the $xy$ plane suggests that the cloud was highly symmetric~\cite{danthesis}. Therefore, we set $\sigma_{TF,y} = \sigma_{TF,z}$ in Eq.~\eqref{eq:AMc} when calculating the amplitude of the condensate monopole mode. Although we observe excitation of other collective modes (dipole and quadrupole), we find that the key features of the experimental results for the monopole mode are independent of whether this assumption is made or $\sigma_{TF,y}$ is simply excluded from Eq.~\eqref{eq:AMc}.

\begin{figure*}[!]
\includegraphics[scale=1]{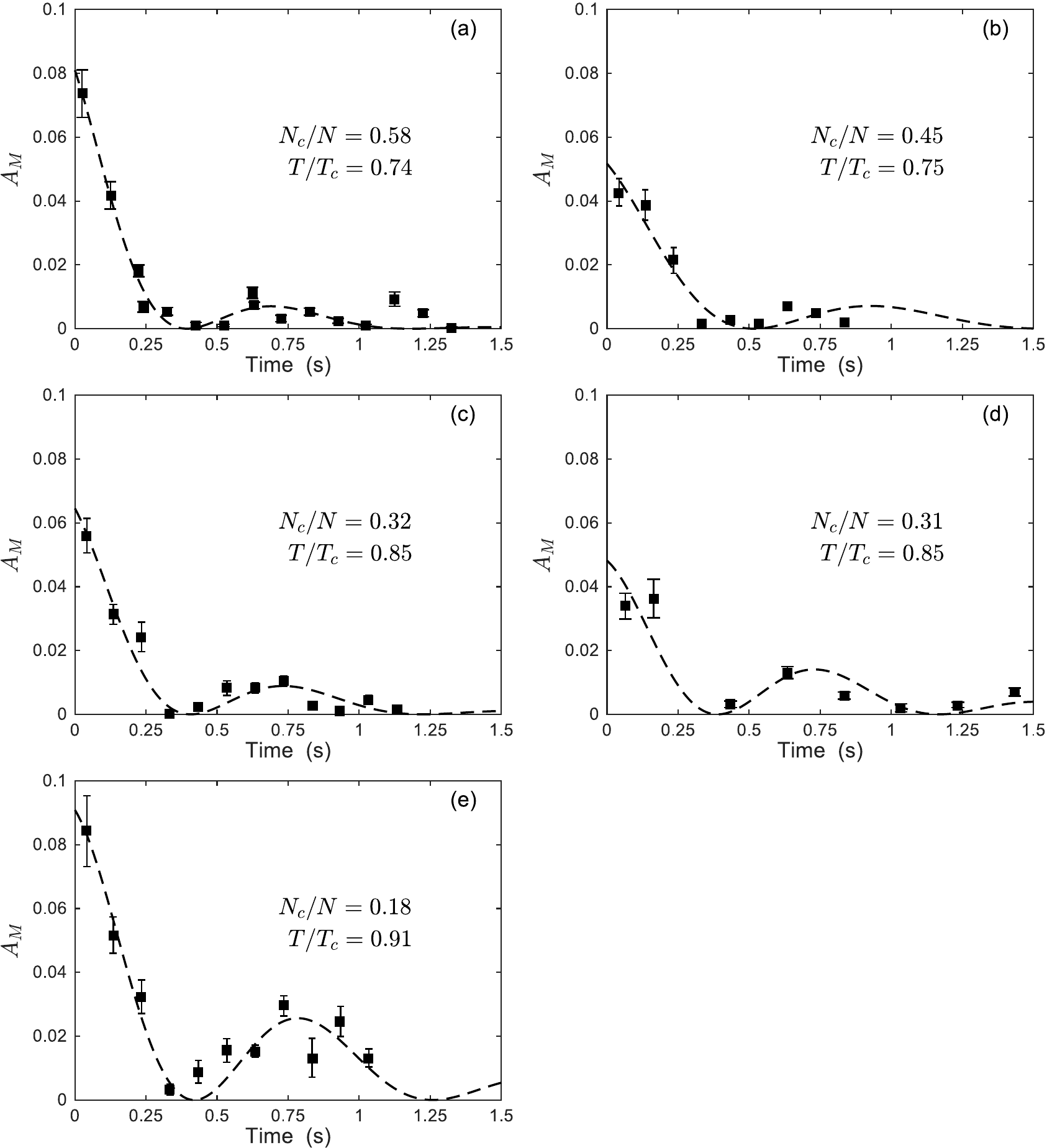}
\caption{\label{fig:expdata1} Amplitude of the monopole mode oscillation (squares) for atom numbers of (a) $N = 8.9\times10^5$, (b) $N = 9.7\times10^5$, (c) $N = 6.7\times10^5$, (d) $N = 5.4\times10^5$, and (e) $N = 7.9\times10^5$. Each frame is labeled with the condensate fraction $(N_c/N)$ and temperature $(T/T_c)$. Error bars represent the statistical uncertainty of multiple realizations of the experiment at each time point. The dashed lines are a fit of the data to Eq.~\eqref{eq:dirty_fit}, which represents the envelope function for the superposition of two sinusoids. From the fit it is found that the beat frequency is $\Delta\omega/\omega_0\sim0.13$ on average and the damping rate lies in the range $\Gamma_\text{e}\sim1.5$--$3.5~\text{s}^{-1}$.}
\end{figure*}

We determine the instantaneous amplitude of the condensate monopole mode by fitting a fixed frequency sine wave to each set of six consecutive time points. The fitting function is of the form
\begin{equation}\label{eq:sine_fit}
g_\sigma(t) = A_\sigma\cos{\left(2\pi\nu t\right)}+B_\sigma\sin{\left(2\pi\nu t\right)}+C_\sigma,
\end{equation}
where $\nu = 19~\text{Hz}$, and $A_\sigma$, $B_\sigma$, and $C_\sigma$ are fit parameters. This functional form is chosen because we are concerned with the amplitude of the mode, not the frequency, which enables a straightforward linear regression analysis for computing $A_\sigma$, $B_\sigma$, and $C_\sigma$. Finally, we present the data in the form of a fractional amplitude given by
\begin{equation}\label{eqn:fadat}
A_M = \frac{A_\sigma^2+B_\sigma^2}{C_\sigma^2},
\end{equation}
where $A_\sigma$, $B_\sigma$, and $C_\sigma$ correspond to the fit parameters of Eq.~\eqref{eq:sine_fit}. The results of this analysis are shown in Fig.~\ref{fig:expdata1}, where time $t = 0$ is defined as the point at which the modulation of the trap frequency ceases.

A central feature of the data is that the amplitude of the condensate monopole mode does not decay exponentially. Across the five data sets, there is a consistent collapse in the amplitude of the monopole mode between $t = 0.3$--$0.5~\text{s}$, and a partial revival around $t = 0.7$--$0.9~\text{s}$. Motivated by the results of the coupled-modes analysis, we fit the data to an envelope function that represents the superposition of two sinusoidal modes, and we include an overall exponential damping factor to represent the loss of amplitude with time:
\begin{equation}\label{eq:dirty_fit}
g_{\text{e}}(t) = A_{\text{e}}\cos{\left(\frac{\Delta\omega}{2}t\right)}^2e^{-\Gamma_{\text{e}} t},
\end{equation}
where $A_{\text{e}}$ is the initial amplitude, $\Delta\omega$ is the beat frequency, and $\Gamma_{\text{e}}$ is the damping rate of the envelope. The results of this fit are overlaid with the data in Fig.~\ref{fig:expdata1}. From the fit, we find $\Delta\omega/\omega_0\sim0.13$ on average, and damping rates in the range $\Gamma_{\text{e}}\sim1.5$--$3.5~\text{s}^{-1}$. The observed beat frequency is less than $\Delta\omega/\omega_0\sim0.2$ as expected from the coupled-modes analysis. We attribute this disagreement to the naive form of the fitting function, which assumes that the two sinusoidal modes damp at the same rate, respond equally to the trap frequency perturbation, and have no phase difference between them. These assumptions are investigated further in the next section where the two sinusoidal modes are identified with the in-phase and out-of-phase modes predicted by the coupled-modes analysis. 

Before moving on, it is important to note that in Ref.~\cite{colussi2015} it was shown that anharmonic corrections to the trap geometry were likely responsible for the anomalous exponential damping of the monopole mode observed above the critical temperature~\cite{breath2015}. As shown in Ref.~\cite{breath2015}, this damping is $<0.2~\text{s}^{-1}$ for clouds with a full width at half maximum (FWHM) of $<125~\mu\text{m}$. In this work, the FWHM of the cloud below the critical temperature satisfies this criterion; thus, we neglect anharmonic corrections to the trap geometry as the observed damping rate is approximately an order of magnitude larger.

\section{Collisional dynamics}\label{sec:zng}
We now investigate the damping observed in the experimental data through numerical simulations within the ZNG formalism. The coupled-modes analysis ignores collisions and exchange of particles between the two components, as well as nonlinear mean-field effects. However, below the critical temperature, these interactions between the condensate and noncondensate can shift the frequencies of collective modes and cause damping. Collisions that exchange energy and particles between the two components of the gas cause collisional damping~\cite{zngbook1,zngbook2}, whereas mean-field effects lead to Landau damping (see Refs.~\cite{landaudamping1, landaudamping2, landaudamping3, landaudamping4, landaudamping5,landauspherical,landauqm,landauzng} for further discussion) and Beliaev damping~\cite{beliaevdamp}. Landau damping describes a process where a collective mode decays due to its interaction with a thermal distribution of excitations, and it is expected to dominate at higher temperatures approaching the critical temperature. On the other hand, Beliaev damping is a process where a collective mode decays into two lower energy excitations, which is suppressed for the lowest energy collective modes of a trapped gas due to the discretization of energy levels. Thus, the Beliaev process is absent for the monopole mode, and is therefore excluded from our analysis in this paper. In the remainder of this section, we discuss the ZNG formalism and describe the results of numerical simulations in the context of the coupled-modes analysis and experimental data already presented.

\subsection{Outline of the ZNG formalism}
The ZNG formalism is a prescription for describing a partially condensed Bose gas by breaking the Bose field operator into a condensed part and a noncondensed part.  It couples a generalized Gross-Pitaevskii equation (GGPE) for the condensate with a QBE for the noncondensate.  It has previously been utilized to study collective oscillations at finite temperature~\cite{quadzng,zngscissors,zng_cmodes1,zngtransbreath}, as well as finite temperature effects on solitons~\cite{zng_soliton}, vortices~\cite{zng_vortex1,zng_vortex2}, and turbulence~\cite{zng_turb}. In addition, recent work by Lee and Proukakis~\cite{lee_preprint} applies the ZNG method to study collective modes, condensate growth, and thermalization dynamics for both single and multicomponent condensates. Here, we outline the basic formalism --- a full description can be found in Ref.~\cite{zngbook2}.

The evolution of the condensate field $\Phi(\textbf{r},t)$ is governed by
\begin{widetext}
\begin{equation}\label{eq:ggpe}
i\hbar\frac{\partial\Phi(\textbf{r},t)}{\partial t} = \left\{-\frac{\hbar^2\nabla^2}{2m}+V(\textbf{r},t)+g[n_c(\textbf{r},t)+2\tilde{n}(\textbf{r},t)]-iR(\textbf{r},t)\right\}\Phi(\textbf{r},t),
\end{equation}
where the number of atoms in the condensate is $N_c = \int d\textbf{r}\left|\Phi(\textbf{r},t)\right|^2$.
In Eq.~\eqref{eq:ggpe}, $m$ is the particle mass, $V(\textbf{r},t)$ is the trapping potential, $g = 4\pi \hbar^2 a_s/m$ is the atom-atom interaction strength with $a_s\approx5.3~\text{nm}$ the $s$-wave scattering length for $^{87}\text{Rb}$, and $n_c(\textbf{r},t) = |\Phi(\textbf{r},t)|^2$ and $\tilde{n}(\textbf{r},t)$ are the density of the condensate and noncondensate, respectively. The non-Hermitian source term $R(\textbf{r},t)$ couples the condensate to the noncondensate as described below.

The noncondensate is represented by the Wigner operator
\begin{equation}\label{eq:wignerop}
\hat{f}({\bf r},{\bf p},t)=\int d{\bf r}'e^{i{\bf p}\cdot{\bf r}'/\hbar}\tilde{\psi}^\dagger\left({\bf r}+\frac{{\bf r}'}{2},t\right)\tilde{\psi}\left({\bf r}-\frac{{\bf r}'}{2},t\right),
\end{equation}
where $\tilde{\psi}$ is the Bose field operator for the noncondensed atoms, ${\bf r}$ and ${\bf r}'$ are the center-of-mass and relative coordinates, respectively, and $\textbf{p}$ is the momentum. Taking the expectation value of the Wigner operator yields the Wigner distribution function $f(\textbf{r},\textbf{p},t)$, which can be shown to obey a QBE~\cite{kadanoffbaym,zngbook1,zngbook2}
\begin{equation}\label{eq:qbe}
\frac{\partial f(\textbf{r},\textbf{p},t)}{\partial t}+\frac{\textbf{p}}{m}\cdot\nabla_{\textbf{r}}f(\textbf{r},\textbf{p},t)-\nabla_{\textbf{r}}U(\textbf{r},t)\cdot\nabla_{\textbf{p}}f(\textbf{r},\textbf{p},t) = C_{12}\left[f,\Phi\right]+C_{22}\left[f\right],
\end{equation}
\end{widetext}
where $U(\textbf{r},t) = V(\textbf{r},t)+2gn_c(\textbf{r},t)+2g\tilde{n}(\textbf{r},t)$ is the effective potential for the noncondensate in the Hartree-Fock approximation, and the right-hand side of this equation describes the effects of interatomic collisions on the distribution function. The density and number of atoms in the noncondensate are defined as
\begin{eqnarray}
\tilde{n}(\textbf{r},t) & = & \int\frac{d\textbf{p}}{(2\pi\hbar)^3}f(\textbf{r},\textbf{p},t), \\
\tilde{N} & = & \int d\textbf{r}~\tilde{n}(\textbf{r},t), \label{eq:Ntilde}
\end{eqnarray}
respectively.

The two collision processes in Eq.~\eqref{eq:qbe} represent a collision between a condensed atom and a noncondensed atom ($C_{12}$) and two noncondensed atoms ($C_{22}$). The former process leads to growth or decay of the condensate, and is the source of the non-Hermitian term in Eq.~\eqref{eq:ggpe},
\begin{equation}\label{eq:Rterm}
R(\textbf{r},t) = \frac{\hbar}{2\left|\Phi\right|^2}\int\frac{d\textbf{p}}{(2\pi\hbar)^3}C_{12}\left[f,\Phi\right],
\end{equation}
where
\begin{eqnarray} \label{eq:c12}
C_{12}\left[f,\Phi\right]&=&\frac{\sigma\left|\Phi\right|^2}{\pi m^2}\int d\textbf{p}_2d\textbf{p}_3d\textbf{p}_4\delta\left(m\textbf{v}_c+\textbf{p}_2-\textbf{p}_3-\textbf{p}_4\right) \nonumber \\
&\times&\delta\left(\epsilon_c+\epsilon_2-\epsilon_3-\epsilon_4\right) \nonumber \\
&\times&\left[\delta\left(\textbf{p}-\textbf{p}_2\right)-\delta\left(\textbf{p}-\textbf{p}_3\right)-\delta\left(\textbf{p}-\textbf{p}_4\right)\right] \nonumber \\
&\times&\left[(1+f_2)f_3f_4-f_2(1+f_3)(1+f_4)\right],
\end{eqnarray}
describes the effect of exchange collisions between the noncondensate and condensate with cross section $\sigma=8\pi a_s^2$.  Binary collisions between noncondensed atoms are represented by
\begin{eqnarray} \label{eq:c22}
C_{22}\left[f\right]&=&\frac{\sigma}{\pi h^3m^2}\int d{\bf p}_2d{\bf p}_3d{\bf p}_4\delta\left({\bf p}+{\bf p}_2-{\bf p}_3-{\bf p}_4\right)\nonumber\\
&\times&\delta\left(\epsilon+\epsilon_2-\epsilon_3-\epsilon_4\right)\left[(1+f)(1+f_2)f_3f_4\right. \nonumber\\
&-&\left.ff_2(1+f_3)(1+f_4)\right].
\end{eqnarray}
In Eqs.~(\ref{eq:c12}) and (\ref{eq:c22}) the delta functions ensure conservation of energy and momentum in a collision, and $f_i$ represents the value of $f(\textbf{r},\textbf{p},t)$ at the phase-space coordinates of particle $i$. The $(1+f_i)$ terms represent Bose enhancement of the scattering process. Furthermore, Eq.~\eqref{eq:c12} depends on the \emph{local} condensate velocity, energy, and chemical potential given by~\cite{zngbook2}
\begin{eqnarray}
\textbf{v}_c & = & \frac{\hbar}{2mi}\frac{\left(\Phi^*\nabla\Phi-\Phi\nabla\Phi^*\right)}{\left|\Phi\right|^2}, \\
\epsilon_c & = & \frac{1}{2}mv_c^2+\mu_c, \\
\mu_c & = & -\frac{\hbar^2}{2m}\frac{\nabla^2\sqrt{n_c}}{\sqrt{n_c}}+V+gn_c+2g\tilde{n},
\end{eqnarray}
where the dependence of these quantities on $\textbf{r}$ and $t$ has been omitted for brevity.

\subsection{Simulation of the experiment}\label{sec:zng_exp}
To model the experiment, we simulate a gas of $N=8\times 10^5$ $^{87}$Rb atoms in a spherically symmetric harmonic trap with $\omega_0 = 2\pi\times9~\text{Hz}$.  Using the algorithm outlined in Ref.~\cite{zngbook2}, we generate equilibrium initial states of the condensate and noncondensate for temperatures ranging from $0.1$--$0.9~T_c$~\footnote{As $T\rightarrow T_c$ the numerical method for calculating the equilibrium state of the gas becomes unstable, and $0.9~T_c$ is an empirical upper bound.}. We then directly simulate the excitation of the monopole mode as in the experiment by sinusoidally modulating the frequency of the trapping potential at $\omega_D=2\omega_0$ for four periods.  We find that our results are essentially unchanged for drive frequencies of $(1+\sqrt{5}/2)\omega_0$ or $\sqrt{5}\omega_0$.  We use trap frequency modulation amplitudes of $\epsilon=0.02$, $0.03$, or $0.04$, and then allow the system to evolve freely for $t = 2~\text{s}$.  We find that this range of $\epsilon$ excites the monopole mode of the condensate with an amplitude comparable to that observed in the experiment (i.e. $10$--$15\%$ peak increase in the TF radius of the condensate from equilibrium). We note that these values are somewhat less than the quoted experimental value of $\epsilon\approx0.1$, and speculate that this discrepancy is a result of multiple collective modes being excited in the experiment due to the difficulty of driving the trap perfectly spherically. This is in contrast to the simulations where only the monopole mode is excited, and therefore less energy is required to be added to the system to achieve the same level of excitation of the condensate.

\begin{figure*}[!]
\includegraphics[scale=1]{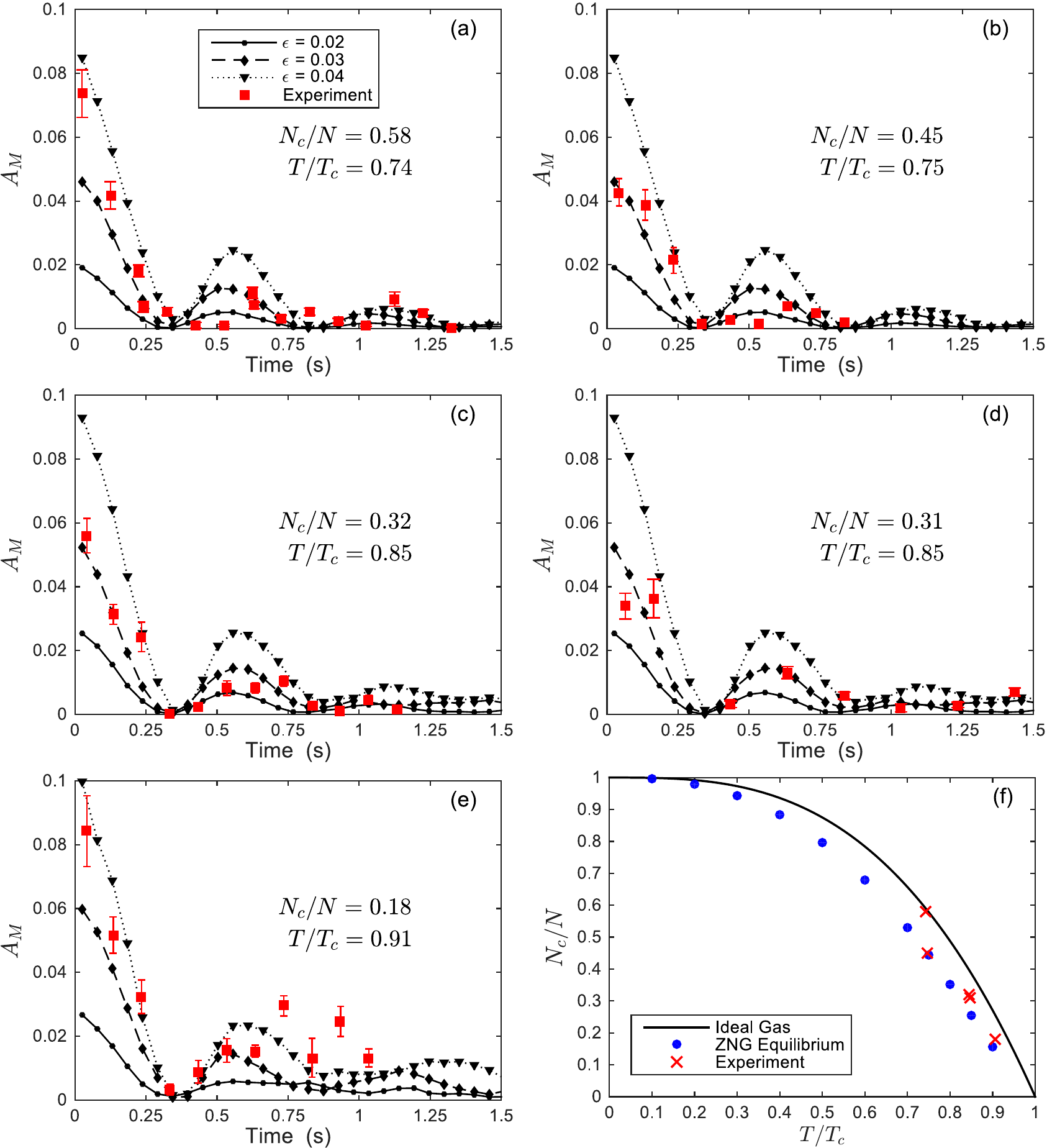}
\caption{\label{fig:expdata2} (Color online) Amplitude of the monopole mode oscillation for experimental atom numbers of (a) $N = 8.9\times10^5$, (b) $N = 9.7\times10^5$, (c) $N = 6.7\times10^5$, (d) $N = 5.4\times10^5$, and (e) $N = 7.9\times10^5$. Each frame is labeled with the condensate fraction $(N_c/N)$ and temperature $(T/T_c)$, and the legend denotes the different modulation amplitudes used in the ZNG simulations. Error bars on the experimental data represent the statistical uncertainty of multiple realizations of the experiment at each time point. (f) Condensed fraction vs temperature for the ideal Bose gas, $N_c/N = 1-(T/T_c)^3$ (solid line), the equilibrium state of the ZNG simulations (blue points), and the experimental data (red crosses). All simulations are performed with $N = 8\times10^5$ atoms.}
\end{figure*}

We record the mean-square radius of the condensate as a function of time, along with snapshots of the individual density profiles. Although the experimental data sets have total atom numbers that range between about $6\times10^5$ and $1\times10^6$, we find simulations for $8\times10^5$ atoms represent the features of interest, namely the collapse and revival behavior and damping rate. In order to compare directly with the experimental data, we generate 2D column densities from the simulation results, and determine the TF radii using the same bimodal fitting routine described in Sec.~\ref{sec:experiment}. Equation~\eqref{eq:AMc} is used to calculate the amplitude of the condensate monopole mode, and Eq.~\eqref{eq:sine_fit} is fit to single periods of the oscillation corresponding to a window of approximately $53~\text{ms}$. The results of this analysis are overlaid with the experimental data in Fig.~\ref{fig:expdata2} for the three different values of $\epsilon$.
The timescale of the first collapse and revival observed in the simulation results show good agreement with the experiment.

\subsection{Extraction of damping rates}\label{sec:zng_analysis}
The prediction of the coupled-modes analysis and results of the ZNG simulations show good agreement with the collapse and revival behavior observed in the experimental data (see Figs.~\ref{fig:expdata1} and~\ref{fig:expdata2}). In addition, the damping observed in the results of the ZNG simulations agrees well with experimental observations. Therefore, due to the limited and noisy experimental data available, we use the results of the ZNG simulations instead of experimental data to get an estimate of the damping rates for the in-phase and out-of-phase eigenmodes predicted by the coupled-modes analysis.

\begin{figure}[!]
\includegraphics[scale=1]{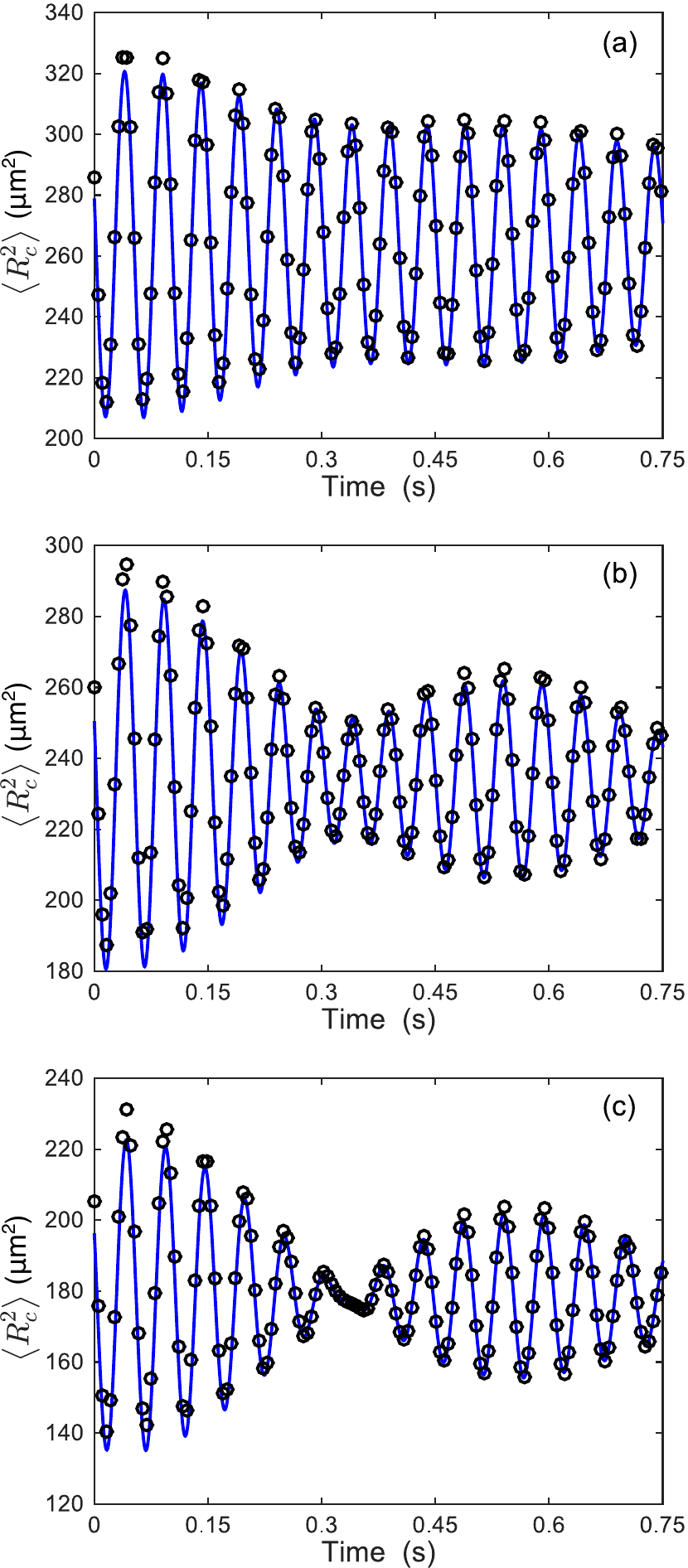}
\caption{\label{fig:rtotfit} (Color online) Simulated mean-square radius of the condensate density for a trap frequency modulation amplitude of $\epsilon = 0.03$ at (a) $T = 0.4~T_c$, (b) $0.6~T_c$, and (c) $0.8~T_c$  (black circles) and resulting fit of Eq.~\eqref{eq:fitfun} (blue line). The density of simulated points has been reduced for clarity.}
\end{figure}

We fit the simulated evolution of the condensate mean-square radius, $\left<R_c^2\right> = \int d\textbf{r}~r^2n_c(\textbf{r})$, by the sum of two sine waves with decaying amplitudes
\begin{eqnarray}\label{eq:fitfun}
g_c\left(t\right) &= &A_1\sin{\left(2\pi\nu_1t+\phi_1\right)}e^{-\Gamma_1t}\nonumber\\
&&+A_2\sin{\left(2\pi\nu_2t+\phi_2\right)}e^{-\Gamma_2t}+C_c,
\end{eqnarray}
where $A_i$, $\nu_i$, $\phi_i$, and $\Gamma_i$ are the amplitudes, frequencies, phases, and damping rates, respectively, of the two eigenmodes, and $C_c$ is a constant offset. Figure~\ref{fig:rtotfit} shows typical results of this fitting procedure for simulation results with a trap frequency modulation amplitude of $\epsilon=0.03$. Time $t = 0$ is defined as the point at which the modulation of the trap frequency ceases. We choose to fit to the mean-square radius of the condensate as its time evolution is most sensitive to the presence of both eigenmodes across the temperature range investigated. The mean-square radius of the total density becomes dominated by the noncondensate at higher temperatures, and any signature of a second eigenmode is lost. Similar behavior is observed in the evolution of the noncondensate mean-square radius.

The mode frequencies extracted from this fitting procedure show excellent agreement with the results of the coupled-modes analysis across the temperature range simulated. Simulation results at temperatures of $T = 0.1~T_c$ and $0.2~T_c$ are fit with a single decaying sinusoid due to the absence of a second mode. This also agrees with the prediction of the coupled-modes analysis, where only a single mode responds to a trap frequency perturbation for $T\leq0.2~T_c$ (see Fig.~\ref{fig:residues}). A notable feature in Fig.~\ref{fig:rtotfit} is a downward shift in the carrier frequency of $\left<R_c^2\right>$ with increasing temperature, an effect also observed in Ref.~\cite{zngbook1} for the monopole mode. As the temperature of the system increases the out-of-phase mode begins to get excited in conjunction with the in-phase mode, and the carrier frequency shifts to a lower frequency because it represents a weighted average of the two independent mode frequencies.

\begin{figure}[!]
\includegraphics[scale=1]{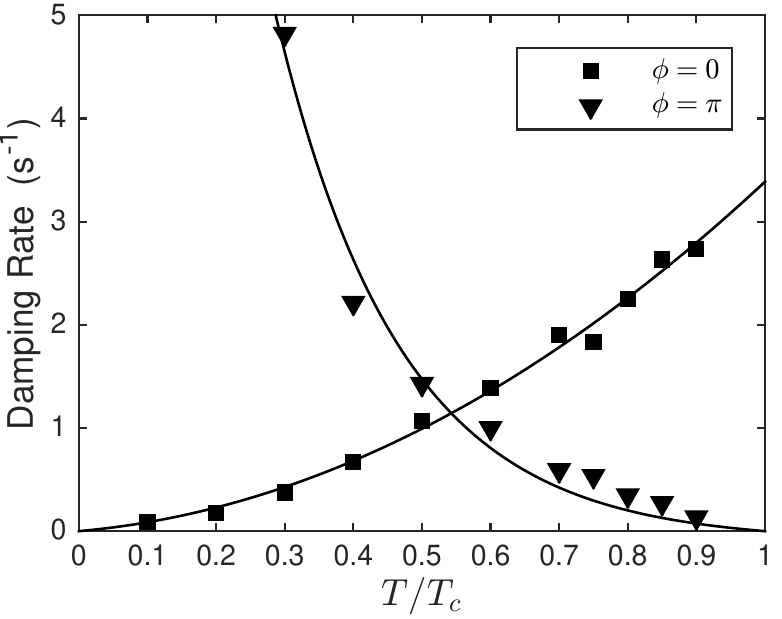}
\caption{\label{fig:damping} Damping rate of the in-phase (squares) and out-of-phase (triangles) mode as a result of fitting Eq.~\eqref{eq:fitfun} to the simulated evolution of $\left<R_c^2\right>$ at each temperature for a trap frequency modulation amplitude of $\epsilon = 0.03$. The solid lines are guides to the eye. Representative fits are shown in Fig.~\ref{fig:rtotfit}.}
\end{figure}

The damping rate of each mode determined from the fitting procedure is shown in Fig.~\ref{fig:damping} as a function of temperature.
In the temperature range of the experiment, there is a mismatch of the damping rates between the two eigenmodes. This mismatch, along with the beating between the two modes, captures the behavior seen experimentally of strong collapse and subsequent revival of the condensate oscillation. At lower temperatures, the in-phase mode dominates and the out-of-phase mode is strongly damped, and the inverse is true at higher temperatures. Thus, the in-phase mode appears to be dominated by the condensate while the out-of-phase mode consists primarily of the noncondensate.

Based on the results of the coupled-modes analysis, one may suspect that the particular drive frequency used to excite the system has a large effect on the nature of the response due to the presence of two resonant excitation frequencies. However, results from ZNG simulations showed little dependence on the drive frequency, which can be attributed to the presence of damping. Damping in the system effectively broadens the resonances such that both modes are appreciably excited when the system is driven in the range $2\omega_0$ to $\sqrt{5}\omega_0$. Thus, the main characteristics of the condensate collapse-revival behavior are relatively insensitive to the particular drive frequency.

As a final note for the interested reader, we have performed additional simulations based on the classical field (c-field) formalism, full details of which are described in Appendix~\ref{appendix:cfield}. For a smaller system of $5\times10^4$ atoms we found reasonable agreement with the ZNG method for the in-phase mode (dominated by the condensate). However, for the out-of-phase mode (dominated by the noncondensate) the two methods show substantial differences in the oscillation frequency and damping rate.

\section{Conclusions}\label{sec:conclusion}
In conclusion, we have experimentally observed non-exponential collapse and subsequent revival of the monopole mode of a finite temperature BEC in an isotropic magnetic trap.  A coupled-modes analysis was used to study the linear response of the system to external perturbation, the results of which are the identification of two eigenmodes of the system corresponding to in-phase and out-of-phase oscillations of the condensate and noncondensate. These modes appear to be collisionless analogs to the first and second sound modes as previously discussed in Ref.~\cite{stoof}. Simultaneous excitation of these two modes results in the observed collapse and partial revival of the condensate monopole mode, which has a timescale compatible with the mismatch in the eigenfrequencies. Damping of the oscillations was also observed experimentally, and simulations within the ZNG formalism resulted in good agreement with the data.

\begin{acknowledgments}
This work was supported by the National Science Foundation Physics Frontier Center (Grant No. PHY1125844). C.J.E.S. and D.Z.A. further acknowledge the support of the Air Force Office of Scientific Research (Grant No. FA9550-14-1-0327). M.J.D. acknowledges the support of the Australia Research Council Discovery Project (No. DP1094025), and the JILA Visiting Fellows program.
\end{acknowledgments}

\appendix
\section{Classical field simulations}\label{appendix:cfield}

An alternative approach to simulate the dynamics of BECs at finite temperature is the classical field, or c-field methodology~\cite{castin2001,blakie2008b,brewczyk_gajda_07}.  Intuitively, a classical field treatment of a BEC will be a reasonable approximation when the system has a significant number of modes with large occupation numbers $ n_i = k_B T /\epsilon_i \gg 1$, where $\epsilon_i$ is the energy of the mode~\cite{blakie_davis_05b}. In these circumstances the GPE can be a good description of the dynamics of this part of the system, rather than only the condensate itself as in the ZNG formalism.  The stochastic projected GPE (SPGPE) methodology~\cite{gardiner_davis_03,stoof_99}  describes the non-classical (or thermal gas) modes as a static finite temperature reservoir, whereas the projected GPE (PGPE) methodology~\cite{davis_morgan_01,Goral2001b,blakie_davis_05a} entirely neglects the coupling to these higher-energy modes.  These techniques have the advantage compared to ZNG that they go beyond the Hartree-Fock treatment of the low-lying modes, and treat their interaction with the condensate nonperturbatively~\cite{blakie2008b}.  However, current implementations of these methods have the significant limitation compared to ZNG in that the majority of the thermal cloud is static (SPGPE) or missing (PGPE), potentially neglecting an important part of the physics when it comes to considering the interplay of the condensate and thermal clouds.  This was pointed out, for example, by Bezett and Blakie~\cite{bezettblakie} who used a PGPE methodology to simulate the experiments of Jin \emph{et al.}~\cite{colmodebec2}. Furthermore, analysis by Karpiuk \emph{et al.}~\cite{karpiuk2010}, who used a similar $c$-field method, found only qualitative agreement with the experiment, emphasizing the importance of thermal cloud dynamics in the behavior of collective modes.

\begin{figure}
\includegraphics[width=8.6cm]{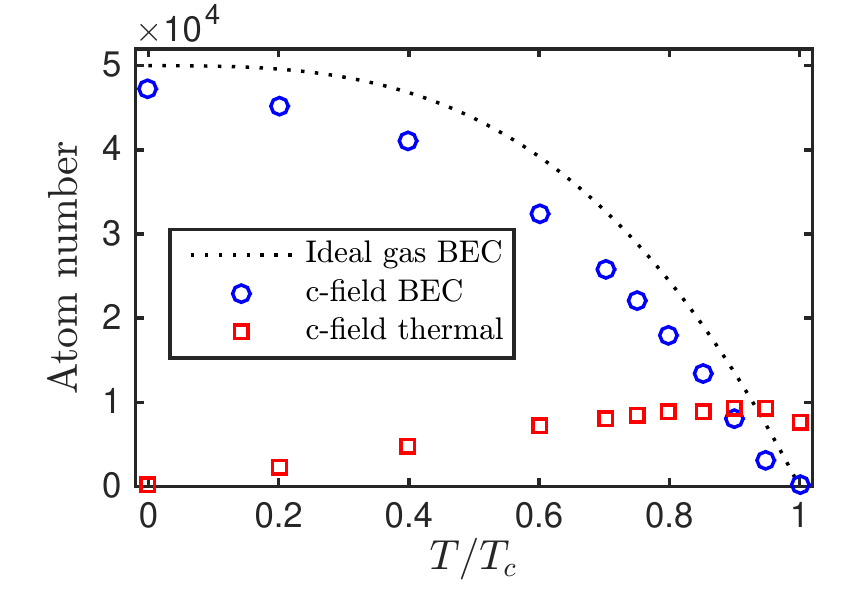}
\caption{(Color online) c-field condensate (blue circles) and thermal cloud (red squares) atom numbers as a function of reduced temperature.  Note that a significant number of thermal cloud atoms with energies above the cutoff are missing at higher temperatures.  The ideal gas condensate number for $N=5\times 10^4$ atoms is shown as a dotted line.}
\label{fig:condfrac}
\end{figure}

Despite these reservations, we have performed c-field simulations of the experiments described in this paper to  gain further insight into the collapse and revival behavior. Due to the unfavorable scaling of harmonically trapped PGPE simulations with atom number~\cite{blakie_08}  we chose to simulate a total of $N=5\times 10^4$ atoms with all other parameters being the same to test the methodology.   Initial states were generated by evolving the SPGPE to equilibrium using ideal gas estimates of the temperature and condensate number.  We chose a range of temperatures from  $T=0$ to $T=T_c \approx 15$ nK, with a corresponding chemical potential estimated from the Thomas-Fermi prediction for the system, $\mu =  (15 a N_0/ a_{ho})^{2/5}\hbar\omega_0/2$, and an energy cutoff of $30 \hbar \omega_0$.  The BEC and thermal atom numbers were determined using time averaging and the Penrose-Onsager criterion~\cite{blakie2008b} as plotted in Fig.~\ref{fig:condfrac}, where it is clear that an increasing number of thermal atoms are missing at higher temperatures.

Following the generation of an initial state, evolution then switched to the number- and energy-conserving PGPE with a range of energy cutoffs from 30--$43 \hbar \omega_0$ to allow for the thermal cloud to respond to being driven.  The results presented below are for the largest of these cutoffs, although there was little quantitative difference between the different values.

The trapping potential was driven for four cycles at three different driving frequencies: $\omega_D = \{2, (1+\sqrt{5}/2),\sqrt{5}\}\omega_0$ with an amplitude of 0.08, i.e., $V(\textbf{r},t) = [1+0.08 \sin(\omega_D t)] V(\textbf{r})$. Again little quantitative difference was found between the results --- those plotted are for $\omega_D = (1+\sqrt{5}/2)$. The system was then allowed to relax in a static trapping potential for 50 trap periods. Ten trajectories were run for each temperature with different equilibrium initial states before averaging over all realizations.  However, for the results presented a single trajectory would suffice to remove almost all visible statistical noise from the results.

\begin{figure}[h]
\includegraphics[width=8.6cm]{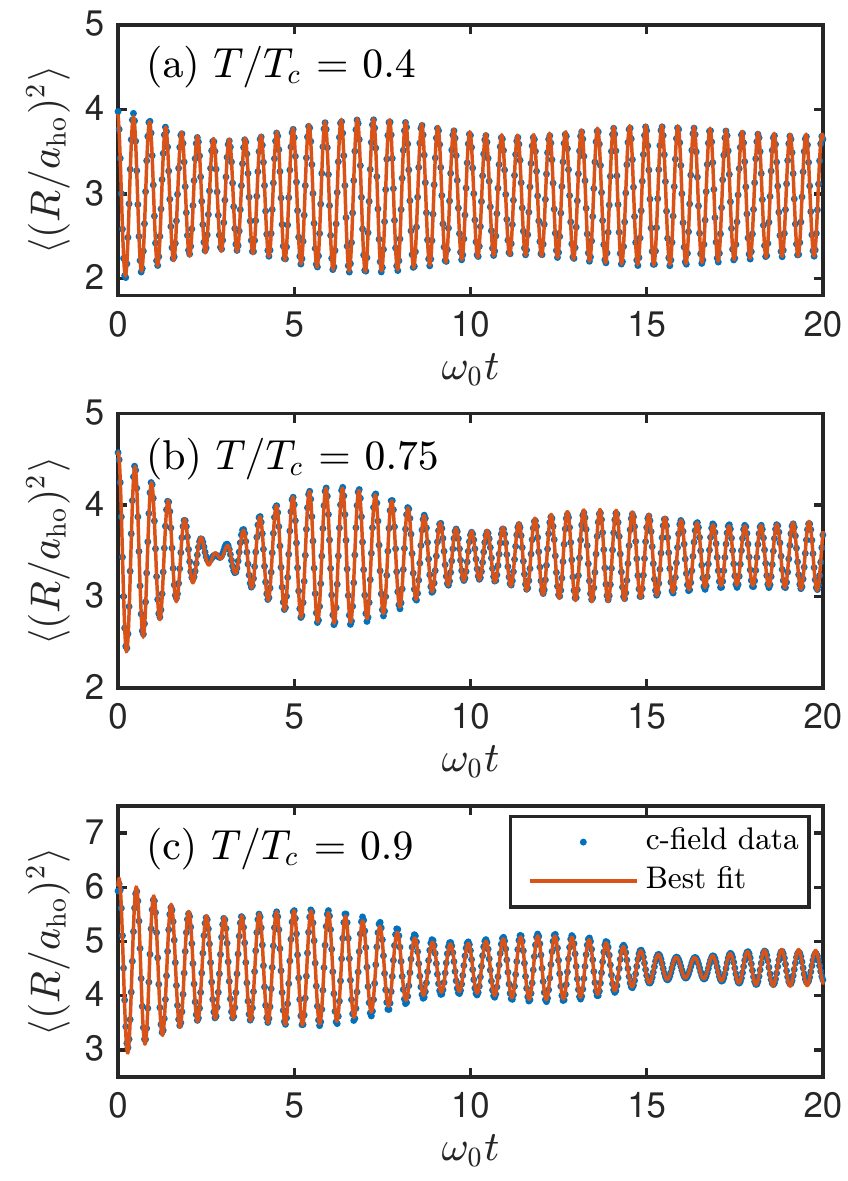}
\caption{(Color online) Relaxation of the monopole mode as indicated by the mean-square radius of the Bose gas from c-field simulations  at three different temperatures: (a) $T/T_c = 0.4$; (b) $T/T_c = 0.75$; (c) $T/T_c = 0.9$.  For these data the driving frequency was $\omega_D = (1+ \sqrt{5}/2) \omega_0$. Similar results are obtained for all other driving frequencies simulated.  Blue dots are results from the simulations, whereas the red lines are a fit to the data from the sum of two exponentially decaying sinusoids.}
\label{fig:fits}
\end{figure}

The observable that provides the most succinct information from these simulations is the expectation value of the mean-square radius of the Bose gas, $\langle R^2 \rangle$.  Sample results are shown in Fig.~\ref{fig:fits} for three different temperatures, where a clear beat signal can be seen --- qualitatively the same behavior as for the ZNG simulations in Fig.~\ref{fig:rtotfit}.  We find that the c-field simulation data is extremely well fit by a the sum of two exponentially decaying sinusoids as in Eq.~\eqref{eq:fitfun}, providing further evidence of two dominant eigenmodes being excited.

\begin{figure}[h]
\includegraphics[width=8.6cm]{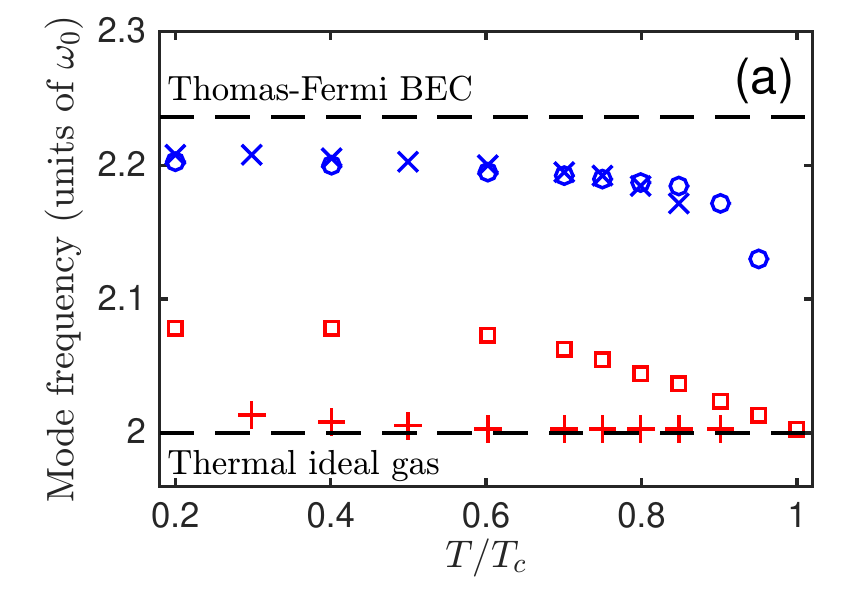}
\includegraphics[width=8.6cm]{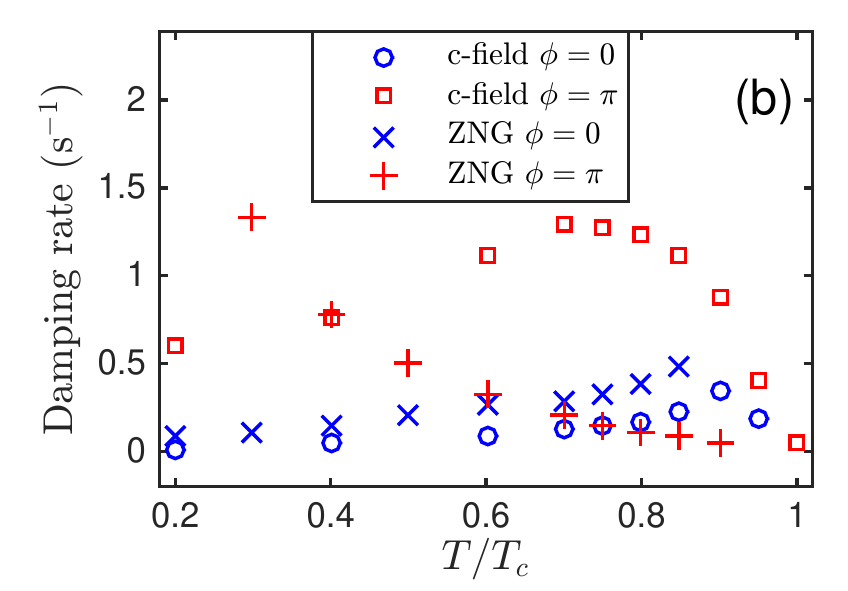}
\caption{(Color online) Comparison of the eigenmode properties for the ZNG and c-field methodologies for $N = 5 \times 10^4$ atoms. (a) Excitation frequencies as a function of the reduced temperature.  Blue circles: condensate, c-field.  Blue crosses: condensate, ZNG.  Red crosses: thermal cloud, c-field.  Red plusses: thermal cloud, ZNG. The hydrodynamic frequency for the monopole mode of the condensate at $T=0$ ($\sqrt{5}\omega_0$), and the expected frequency for a thermal cloud above $T_c$ ($2\omega_0$) are both indicated as horizontal dashed lines. (b) Decay rates as a function of the reduced temperature.  The legend is the same as for (a).}
\label{fig:freq}
\end{figure}

The extracted mode frequencies and damping rates are shown in Fig.~\ref{fig:freq}(a) and Fig.~\ref{fig:freq}(b), respectively, with a comparison to those from corresponding ZNG simulations.   In Fig.~\ref{fig:freq}(a) the eigenmode frequencies for the $\phi = 0$ mode closest to the hydrodynamic result for a Thomas-Fermi condensate are somewhat below  $\sqrt{5}\omega_0$ due to the system not being in the deep Thomas-Fermi regime.  However, the c-field and ZNG frequencies are in good agreement. The c-field out-of-phase mode frequency is quite a bit higher than $2\omega_0$ at the lowest temperatures, presumably due to being dominated by the condensate.  However, it does tend toward $2 \omega_0$ as the condensate gets smaller for $T\rightarrow T_c$.

The damping rates shown in  Fig.~\ref{fig:freq}(b) show some discrepancy between the two methodologies.  While the in-phase modes are in reasonable agreement, the out-of-phase mode damping rates are quite different.  We believe that this is due to the fact that only a  fraction of the thermal cloud is described by the c-field simulations, and the mean-field interactions of the c-field density and the missing thermal component is an important ingredient in correctly describing the dynamics.  In principle the c-field methodology could be extended such that the thermal cloud is described by a quantum Boltzmann equation as in the ZNG approach, however this has yet to be implemented. Our conclusion is that the ZNG methodology currently offers the most complete description of collective oscillations in condensed Bose gases at finite temperature.

\section{Numerical solution of the spherically symmetric ZNG equations}\label{appendix:zng}
The ZNG equations describe the evolution of a degenerate Bose gas in a six-dimensional phase space $(\mathbf{r},\mathbf{p})$, which is a computationally demanding problem. However, making use of symmetries to reduce the number of degrees of freedom needed to describe the system can lead to significant numerical advantages. Here, we make use of the spherical symmetry of the trap which reduces the dimensionality of the problem from six to three, leaving a radial displacement $r$, a momentum magnitude $p$, and an angular variable, $\cos{\theta}$, describing the orientation of the vector $\mathbf{p}$ with respect to $\mathbf{r}$. The main algorithm for numerically solving the ZNG equations is discussed in detail in Refs.~\cite{zngbook1,zngbook2}, and in this Appendix we describe the additional details needed to apply this algorithm to solve the GGPE and QBE in a spherically symmetric geometry.

With spherical symmetry the condensate wavefunction depends only on $r$, and a 1D GGPE can be used to describe its evolution. Furthermore, rewriting the GGPE in terms of the variable $\phi(r) = r\Phi(r)$ eliminates the first derivative term in the Laplacian, allowing for application of simple Dirichlet boundary conditions where $\phi(r)\rightarrow0$ as $r\rightarrow0,\infty$. We employ the Crank-Nicolson method~\cite{hollandCN} to solve the GGPE in this form.

\begin{figure}[!]
\includegraphics[scale=0.65]{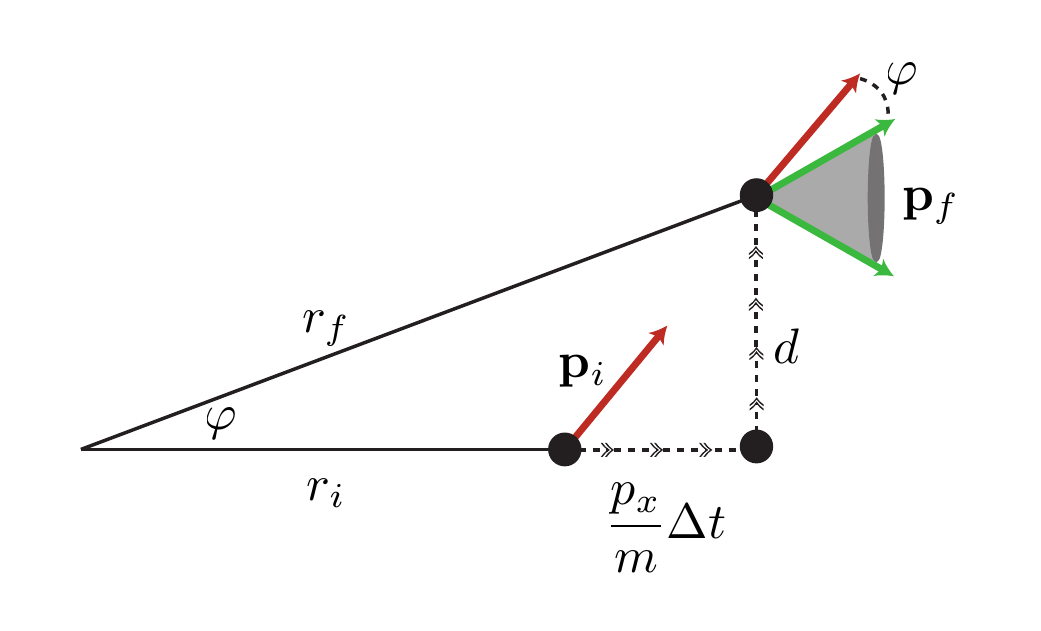}
\caption{\label{fig:particle_mover} (Color online) The particle is rotated from its initial trajectory along $r_i$ by an angle $\varphi$ due to the off axis components of $\textbf{p}_i$. After the particle position is updated to $r_f$ the momentum components are rotated and realigned with the position vector. Due to spherical symmetry the azimuthal angle is not unique and is chosen randomly. This is represented by the area of revolution of $\textbf{p}_f$ about the final position vector $r_f$.}
\end{figure}

As in Ref.~\cite{zngbook1}, a tracer particle method is used to evolve the noncondensate distribution function in phase space such that a Monte Carlo sampling method can be employed to simulate the effects of collisions. We use $2\times10^5$ tracer particles for all simulations presented in this paper. At each time step the tracer particle positions and momenta are updated based on Newton's equations of motion. Following the method outlined by Bird~\cite{birdbook}, we take advantage of the spherical symmetry by only storing the radial coordinate of each tracer particle. However, the complete motion of each particle in 3D space must be tracked such that three momentum components are stored for each particle. At the beginning of each time step we utilize the rotational symmetry of the problem and arbitrarily align the position vector of each particle with the $x$ axis. The action of the $y$ and $z$ directed momentum components is to then push the particle off this axis. It is straightforward to calculate the new radial position of the particle; however, the off-axis motion causes a rotation of the particle trajectory and the momentum components must be rotated accordingly. The new particle position on the $x$ axis is
\begin{equation}
x = r_i+\frac{p_x}{m}\Delta t,
\end{equation}
where $r_i$ is the initial radial position of the particle, $p_x$ its momentum along the $x$ axis, and $\Delta t$ is the length of the current time step. The action of $p_y$ and $p_z$ moves the particle off axis by a distance
\begin{equation}
d = \sqrt{\left(\frac{p_y}{m}\Delta t\right)^2+\left(\frac{p_z}{m}\Delta t\right)^2},
\end{equation}
such that the new radial position $r_f$ of the particle is
\begin{equation}
r_f = \sqrt{x^2+d^2}.
\end{equation}
The sine and cosine of the rotation angle are then given by
\begin{eqnarray}
\sin{\varphi} = d/r_f, \\
\cos{\varphi} = x/r_f,
\end{eqnarray}
and an azimuthal angle is chosen at random such that $\phi\in\left[0,2\pi\right]$. Finally, the new momentum components are calculated,
\begin{eqnarray}
p_{x,f} & = & p_x\cos{\varphi}+\sqrt{p_y^2+p_z^2}\sin{\varphi}, \\
p_{y,f} & = & p_{c,f}\sin{\phi}, \\
p_{z,f} & = & p_{c,f}\cos{\phi},
\end{eqnarray}
where $p_{c,f} = -p_x\sin{\varphi}+\sqrt{p_y^2+p_z^2}\cos{\varphi}$. Figure~\ref{fig:particle_mover} provides a graphical representation of the particle movement algorithm.
Note that although three momentum components are stored for each particle in addition to the position, the algorithm is effectively three-dimensional since the azimuthal angle is randomized at each time step.

\begin{figure}[!]
\includegraphics[scale=0.65]{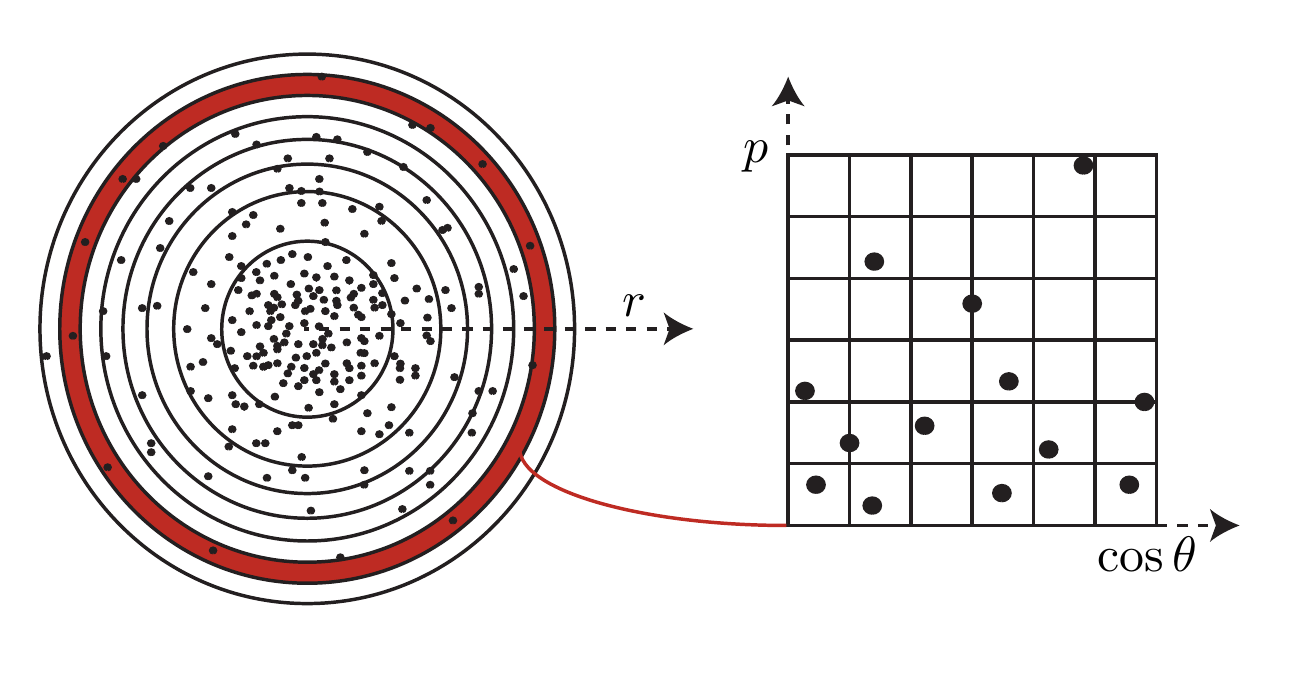}
\caption{\label{fig:binning} (Color online) Cartoon depiction of the phase space binning process. Particles (black points) are binned in position space using constant volume shells with the radial width of each bin decreasing with $r$. Within each spatial bin the particles are further binned in momentum space using a 2D grid of equal area bins based on the magnitude of their momentum $p$ and trajectory $\cos{\theta} = \hat{\textbf{p}}\cdot\hat{\textbf{r}}$.}
\end{figure}

After the tracer particles are moved they are binned in phase space to get an estimate of the local noncondensate density and collision rates. The particles are first binned in radial shells using a constant volume binning scheme. Given the size of the simulation domain, $l_r$, and the total number of bins, $N_b$, the position of each bin edge is given by
\begin{equation}
r_{b,i} = l_r\left(\frac{i}{N_b}\right)^{1/3},
\end{equation}
where $i\in\left[0,N_b\right]$ is an integer representing the bin index. The simulations performed here use $l_r = 60a_{ho}$ and $N_b = 8\times10^4$ where $a_{ho} = \sqrt{\hbar/m\omega_0}$ is the harmonic-oscillator length. This scheme results in wider bins near the origin, and progressively narrower bins as $r$ increases, which we find  reproduces the equilibrium collision rates more accurately than a scheme with equal width bins in $r$. Once the particles are binned in space a 2D scheme is implemented for binning the particles in momentum space based on $p$ and $\cos{\theta}$, where $p$ is the magnitude of the particle momentum. The momentum space bins are equally spaced, and we use $20$ bins in $p$ and $10$ bins in $\cos{\theta}$ for the simulations performed here. Figure~\ref{fig:binning} provides a graphical representation of the binning procedure.

After binning, the tracer particles are used to reconstruct the noncondensate density and phase space distribution function on the discrete numerical grid defined for evolution of the condensate. The density term is required for updating the condensate wave function as well as the momentum of the tracer particles, whereas the phase-space distribution function is necessary for computing the collision rates. Typically, a cloud-in-cell method~\cite{PICbook} is employed to reconstruct a discrete function (e.g.,~density) from the tracer particle distribution by linearly weighting each particle to the nearest grid points defined by the binning process (i.e.,~the edges of each bin). Following this weighting step, the reconstructed function can be interpolated from the binning grid to another numerical grid if necessary. However, in spherical coordinates a linear weighting scheme results in errors, particularly near the grid boundaries~\cite{radialPIC}. Therefore, we employ a volume weighting scheme where the particles are weighted to grid points in proportion to the volume of space between the particle and a given grid point~\cite{splinePIC,volumePIC}. We find this technique improves the accuracy of function reconstruction from the tracer particle distribution, thus allowing fewer tracer particles to be used, which results in improved computation speed.

\end{document}